# EXPERIMENTAL EVIDENCE FOR ELECTROWEAK CORRECTIONS BEYOND FERMION LOOPS[†‡]

Dieter Schildknecht

University of Bielefeld, Department of Theoretical Physics,

33501 Bielefeld, Germany

## Abstract

We reemphasize the importance of discriminating fermion-loop and bosonic electroweak corrections in the analysis of electroweak precision data. Most recent data are indeed precise enough to require corrections beyond (trivial) fermion loops. An analysis of these data in terms of the observables $\Delta x \equiv \epsilon_{N1} - \epsilon_{N2}$, $\Delta y \equiv -\epsilon_{N2}$ and $\epsilon \equiv -\epsilon_{N3}$ identifies the required additional corrections as vertex corrections at the $W^\pm f\bar{f}'$ and $Z^0 f\bar{f}$ vertices. Standard-model values for these corrections are consistent with the experimental data.

[†]Supported by the Bundesminister für Forschung und Technologie, Bonn, Germany.

[‡]Presented at the XXIXth Rencontres de Moriond, March 1994.

This is a brief report on recent theoretical results on the significance of electroweak precision tests obtained by the collaboration of the authors in ref.[1]. The results refine and expand our previous work [2, 3] on this subject.

As stressed a long time ago [4] by Gounaris and myself, in the analysis of electroweak precision tests, it is essential to clearly discriminate between two sources of electroweak one-loop corrections, *fermion-loop* (vacuum polarization) corrections to $\gamma$, $W^\pm$ and $Z^\circ$ propagation on the one hand, and *bosonic vacuum polarization and vertex corrections* on the other hand. The reason for the importance of such a discrimination is obvious. The properties of the (light) *fermions* are *empirically well-known* and the mentioned fermion-loop corrections can accordingly be calculated precisely and uniquely upon introducing the mass of the top quark, $m_t$, as a free parameter. In contrast, the additional bosonic corrections contain trilinear and quadrilinear couplings among the vector bosons and to the Higgs scalar which are *empirically entirely unknown*. The difference between the fermion-loop calculations and the full one-loop standard model results thus sets the scale [4] for the accuracy to be aimed at with respect to genuine quantitative experimental tests of the electroweak theory beyond fermion-loops.

In figs. 1-3, we show the three projections of the three-dimensional 68% C.L. volume defined by the data in $(M_{W^\pm}/M_Z, \bar{s}_W^2, \Gamma_l)$-space in comparison with various theoretical results. The data represent the most recent results from the four LEP collaborations, from SLD and from CDF/UA2 presented at this conference [5], $M_Z = 91.1899 \pm 0.0044 GeV$, $M_{W^\pm}/M_Z = 0.8814 \pm 0.0021$, $\Gamma_l = 83.98 \pm 0.18 MeV$, $\bar{s}_W^2$ (all asymmetries LEP) $= 0.23223 \pm 0.00050$, and $\bar{s}_W^2$ (all asymmetries LEP + SLD) $= 0.23158 \pm 0.00045$.

The theoretical results shown in figs. 1-3 are as follows,

(i) the $\alpha(M_Z^2)$ tree-level prediction (denoted by a star) based on $\alpha(M_Z^2) = 1/128.87 \pm 0.12$ [6] which takes into account the change in $\alpha$ from $\alpha(0)$ to $\alpha(M_Z^2)$ due to lepton and quark loops,

(ii) the full fermion-loop prediction, which takes into account the full contribution of all leptons and quarks to the $\gamma$, $W^\pm$ and $Z^\circ$ propagators, the mass $m_t$ being varied in steps of 20 GeV (and indicated by squares),

(iii) the full standard $SU(2)_L \times U(1)$ one-loop predictions for Higgs masses of $m_H = 100$ GeV (solid line), 300 GeV (long-dashed line) and 1000 GeV (short-dashed line), 20 GeV steps indicated by circles.

From figs. 1-3, we conclude that the present high-precision data deviate from the $\alpha(M_Z^2)$

tree-level prediction and from the full fermion-loop results. The data are accurate enough to require additional contributions beyond fermion loops, and such contributions are indeed provided by the standard bosonic corrections. A top mass of $m_t \simeq 160$ GeV is required for consistency between experiment and standard theory.

The results in figs. 1-3 can be illuminated by an analysis in terms of the parameters $\Delta x, \Delta y$ and $\epsilon$ which within the framework of an effective electroweak Lagrangian [1, 2] specify possible sources of $SU(2)$ violation[1]. The parameters $\Delta x, \Delta y$ and $\epsilon$ can be deduced from the experimental data on $M_{W^\pm}/M_Z, \bar{s}_W^2$ and $\Gamma_l$ and compared with standard one-loop results.

The results for $\Delta x, \Delta y, \epsilon$ thus obtained are displayed in figs. 4-6. The results are striking. According to fig.4, the fermion-loop predictions for $\Delta x$ and $\epsilon$ practically coincide with the complete one loop results, $\Delta x \simeq \Delta x(\text{fermion loops}), \quad \epsilon \simeq \epsilon(\text{fermion loops})$, i.e., the $m_H$-dependent standard bosonic vacuum-polarization effects in $\Delta x$ and $\epsilon$ are of minor importance (and vanishingly small for large values of $m_H$). In contrast, in figs. 5 and 6, we find a significant non-fermion-loop contribution to $\Delta y$, $\Delta y \simeq \Delta y(\text{fermion loops}) + \Delta y(W^\pm-\text{vertex plus box}, Z^0-\text{vertex})$, which is due to vertex (and box) corrections to the $W^\pm f\bar{f}'$ vertex (entering the analysis via the Fermi coupling $G_\mu$ extracted from $\mu$ decay) in conjunction with $Z^0 f\bar{f}$ vertex corrections. *The differences between fermion-loop and full-one-loop theoretical results in figs. 1-3 accordingly have been traced back to significant genuine electroweak $W^\pm f\bar{f}'$ vertex (and box) corrections appearing in conjunction with $Z^0 f\bar{f}$ vertex corrections in the parameter $\Delta y$.*

It is remarkable that the experimental data have *reached a precision which allows one to isolate loop corrections beyond fermion loops*. More specifically, the data require significant vertex corrections. The magnitude of the required corrections is consistent with the prediction of the standard electroweak theory.

---

[1] The parameter $\Delta x$ quantifies global $SU(2)$ violation via $M_{W^\pm}^2 \equiv (1+\Delta x)M_{W^0}^2$, while $\Delta y$ and $\epsilon$ quantify $SU(2)_L$ violation in vector-boson couplings to fermions (with $g_{W^\pm}^2(0) \equiv 4\sqrt{2}G_\mu M_{W^\pm}^2$), namely $g_{W^\pm}^2(0) \equiv (1+\Delta y)g_{W^0}^2(M_Z^2)$, and via mixing $L_\text{mix} \equiv (e(M_Z^2)/g_{W^0}(M_Z^2))(1-\epsilon)A_{\mu\nu}W_3^{\mu\nu}$. The parameters $\Delta x, \Delta y, \epsilon$ are related to the parameters of Altarelli et al [7] via $\Delta x = \epsilon_{N1} - \epsilon_{N2}, \Delta y = -\epsilon_{N2}, \epsilon = -\epsilon_{N3}$.

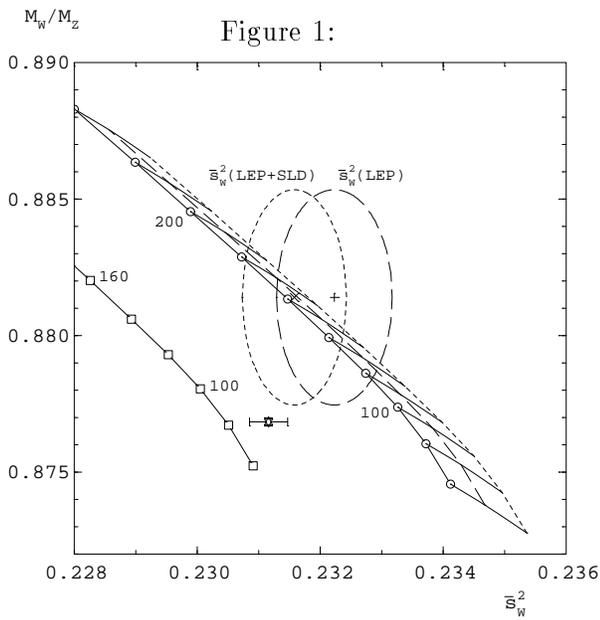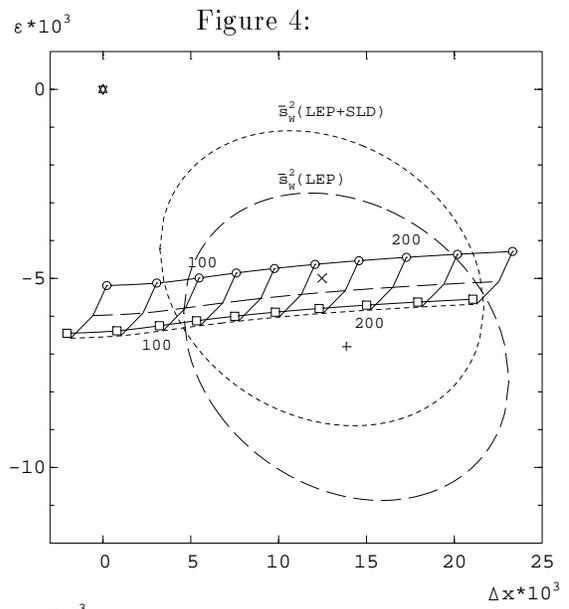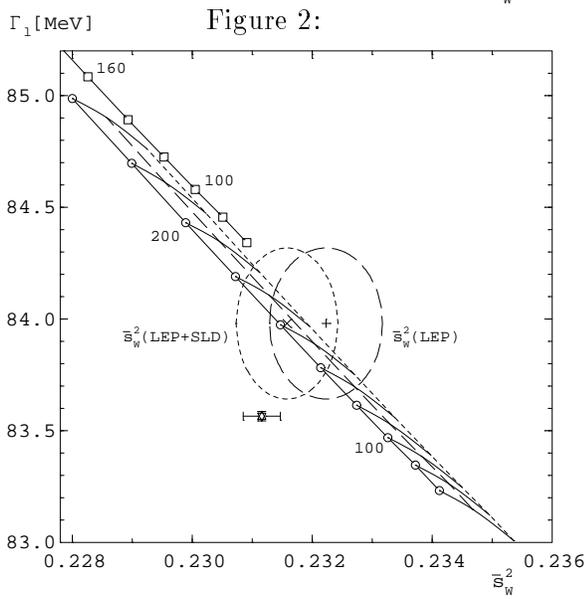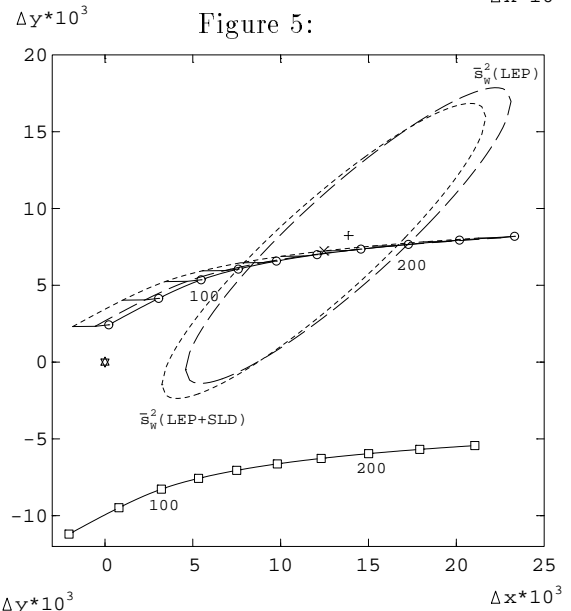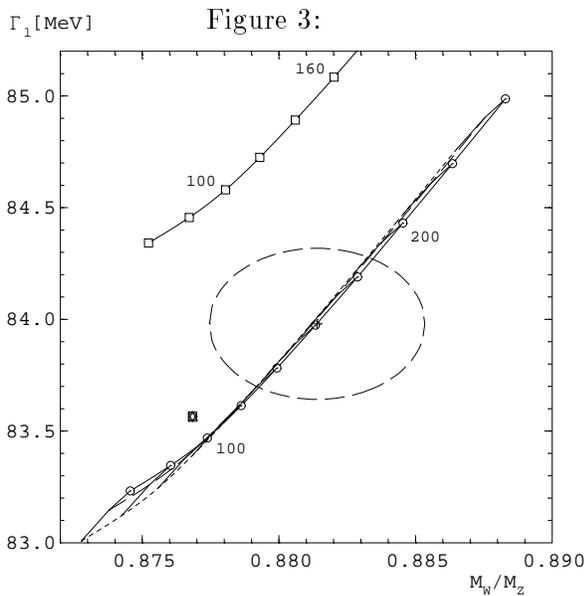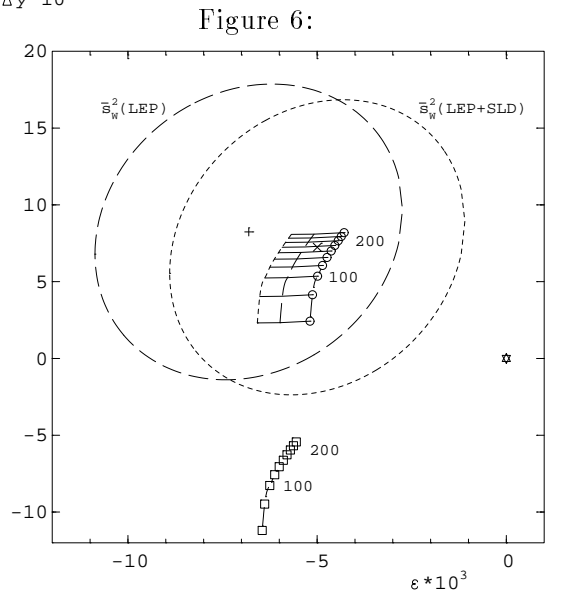

Figs. 1,2,3: The experimental data on $(M_W/M_Z, \bar{s}_W^2, \Gamma_l)$ compared with theory.

Figs. 4,5,6: The experimental data on $\Delta x, \Delta y, \epsilon$ compared with theory.